# The interface heterogeneous nucleation and dynamic dispersion of nuclei in solidification process

Xiaoping Ma[a]

Shenyang National Laboratory for Materials Science, Institute of Metal Research, Chinese Academy of Sciences, Shenyang 110016, China

By comparing the grain sizes under different nucleation conditions, the different nucleation mechanisms were investigated. The primitive nuclei origin at some specific interface, and subsequently disperse into the bulk melt with melt flow. The survival probability of nuclei decides the nuclei density in bulk melt, and the growth of survived nuclei finally constitutes the solidified grains. The nucleation process is highly dynamic, evolving and variable with actual experimental condition and production condition.

## 1. Introduction

For the metal materials, the solidified microstructures will dramatically influence the mechanical properties. In the long history of international solidification research field, the refined equiaxed grain in solidification is always one of the focuses. In order to acquire the refined solidified grains, high nuclei density should be provided. Thus, it is crucial to figure out the nucleation theory and clarify the origin and process of nucleation. Generally, it is considered that the nuclei form at where the bulk melt is

Tel: 86-024-83970106, e-mail: xpma@imr.ac.cn.

undercooled. And it has been realized that the heterogeneous nucleation is the dominant mechanism in practical solidification process of metal and alloy [1-4]. Therefore, effective nucleant particles arouse much research [5-8]. Further, the constitutional undercooling caused by the segregation around the nucleant particles is considered as a contributing factor for further nucleation in front of the crystal/melt interface [8, 9]. Greer further proposed that the nucleation stage is not the controlling factor for the final grain size. The number of grains is determined by a free-growth condition in which a grain grows from a refiner particle at an undercooling inversely proportional to the particle diameter [10]. Based on above understandings, some nucleation model is further applied in the subsequent solidification phenomena research, such as the columnar to equiaxed grains transition or macrosegregation simulation [11-15]. Another important nucleation mechanism weighs the dendrite fracture and nuclei proliferation [16-18], which is widely accepted and applied in explaining some research results about grains refinement through natural convection, mechanical or electromagnetic stirring process [19, 20].

It should be noted that all above works only consider the nuclei originated in the bulk of the melt. In fact, the nucleation process may be more complicated. Solthin [21] has proposed that the nuclei will form at the melt free surface. And such nuclei will settle into the bulk melt. Ohno [22, 23] has proposed that the nuclei will form at the mold wall. And such nuclei will move into the bulk melt with the melt flow. Although Solthin and Ohno's nucleation mechanisms are considered as contribution factors for nuclei density, the contribution proportion is obscure and disappreciated.

Easton and StJohn also suggested that grains can form at the wall of the casting and survive as they move into the bulk of the casting [24]. In my previous works, I also found that the grain refinement is attributed to the nucleation and dispersion from the mold wall [25-28]. These results hint that the nuclei mechanism may be also contributed by the interface heterogeneous nucleation and dynamic nuclei dispersion, which is not fully realized. Thus, in this article, the interface heterogeneous nucleation and dynamic dispersion is further researched and its contribution to nuclei density is evaluated qualitatively.

2. **Experimental procedures**

In order to exclude the grain size evolution in solid phase transition and reflect the solidification grain size, the Al-5%Cu alloy is chosen as the experiment alloy, which has no solid phase transition after solidification. The Al-5%Cu alloy was melted at 950℃ and was hold for 20 minutes in a resistance furnace. Then, it was poured into a graphite crucible (Φ60×60mm, preheated at 660℃). In order to clarify the interface heterogeneous nucleation and dynamic dispersion, a serious of samples with stainless steel sieves (mesh 200), pulsed magnetic field (produced by 5Hz/100V exciting capacitor) and periodic melt surface penetration with a Φ3mm alumina tube (one time per 2 seconds with the penetration depth 10mm below the melt surface) were prepared. The experimental apparatus is shown in Figure 1. After the ingots solidified, the ingots were sectioned, grinded, polished and etched with a $HNO_3$-HCl-HF-$H_2O$ solution. The solidified macrostructures and grains were observed.

## 3. Experimental results

In the base experiment, a sieve barrel was placed in the graphite crucible in order to isolates the center melt and the surrounding melt near the mold wall. The solidified macrostructure is shown in Figure 2(a). It can be seen that coarse equiaxed dendrites are produced in the inner of the sieve barrel, while coarse columnar dendrites are produced outside of the sieve barrel. The average grain diameter in the inner of the sieve barrel is 9.2 mm.

In experiment 1, the pulsed magnetic field was applied after pouring process based on the experiment condition of base experiment. The solidified macrostructure is shown in Figure 2(b). It can be seen that refined equiaxed dendrites are produced both in the inner and outside of the sieve barrel. The average grain diameter in the inner of the sieve barrel is 2.4 mm.

In experiment 2, the nucleation excitation by periodic alumina tube penetration is applied after pouring based on the experiment condition of base experiment. The solidified macrostructure is shown in Figure 3(a). It can be seen that refined equiaxed dendrites are produced in the inner of the sieve barrel, while coarse columnar dendrites are produced outside of the sieve barrel. It should be noted that nonuniform grain size appears in the center equiaxed grain zone. The finer grains with average grain diameter about 98 μm appear at the position below and surrounding the alumina tube penetration position, as shown in Figure 3(b). The coarser grains with average grain diameter about 503 μm, as shown in Figure 3(c), surround the finer grains.

In experiment 3, both the pulsed magnetic field and the nucleation excitation through periodic alumina tube penetration are applied after pouring process based on the base experimental condition. And an additional sieve further isolates the center melt into an upper part and a lower part. The solidified macrostructure is shown in Figure 4. It can be seen that refined equiaxed dendrites with average grain diameter about 1.5 mm are produced in the upper part of the inner melt, while coarse equiaxed dendrites with average grain diameter about 5.1 mm are produced in the lower part of the inner melt.

**4. Discussions**

There is no solid phase transition for the Al-5%Cu alloy after solidification, and the coarsening or ripening of solidified grains are negligible for my sample dimension and cooling condition. Therefore, the final observed grain size can reflect the grain size just after solidification. For this experimental alloy, nucleant particles, such as $TiB_2$ or $Ti_3Al$, are absent. Thus, the condition for the growth restriction model [10] is also absent. The solidified grains size should has no relation with the crystal growth process. Therefore, I can reasonably judge the nuclei density according to the observed grain size. The finer grains coincide with the higher nuclei density. The preliminary qualitative evaluation about the origin mechanism of nuclei is valuable.

4.1 Capability of heterogeneous nucleation in bulk melt or dendrite fracture

In the base experiment, the high mold preheat temperature will eliminate the undercooling condition and nucleation at the mold surface in the pouring process. When the temperature of the mold decreases to the needed temperature for nucleation, the intense melt convection caused by pouring process has ceased. Thus, the nuclei have little chance to detach from the mold wall. The sieve barrel placed in the graphite crucible will further eliminate the nuclei dispersion from the mold wall into the inner melt.

In the base experiment, the stainless steel sieve (mesh 200) is very thin and it was preheated at 660℃. Especially, the melt is poured into the mold at 950℃. The melt will quickly heat the thin stainless steel sieve to the temperature above the liquidus temperature. Thus, nucleation will not happen on the stainless steel sieve.

Similarly, the high pouring temperature will depress the nucleation at the melt surface. When the temperature of the melt surface decreases to the needed temperature for nucleation, the melt convection caused by pouring is weak. Thus, a solidified shell will form on the melt surface. Only a few nuclei may settle from the melt surface before the formation of surface shell.

In this situation, even if we attribute all the nucleation to heterogeneous nucleation in bulk melt or dendrite fracture, the nucleation capability and nuclei density is poor, as shown in Figure 2(a) about the solidified macrostructure. It can be seen that coarse equiaxed dendrites are produced in the inner of the sieve barrel.

4.2 Interface heterogeneous nucleation and dynamic dispersion: originate from oxide film

After the melt is poured into the mold, an oxide film will from on the melt surface. It is reasonable that the oxide film provides the most significant undercooling and favorable nucleation positions. In the base experiment, the primitive nuclei grow to form a solid shell rapidly, and fewer nuclei disperse into the bulk melt. While the pulsed magnetic field produces the melt vibration and disturbs the rapid formation of solid shell, which increases the nuclei dispersion into the bulk melt from the melt surface. Because the high mold temperature and sieve barrel has excluded the influence of nuclei detachment from the mold wall, it can be safely concluded that the refine effect with pulsed magnetic field in the upper part of inter melt in experiment 3 and in the inter part in experiment 1 is caused by the considerable nuclei dispersion originated from the melt surface. In the experiment 3, the horizontal sieve can prevent the dispersion on nuclei into the lower part, therefore distinct grain sizes are observed. The distinct grain size difference between the upper and lower part in experiment 3 further affirms that the heterogeneous nucleation in the bulk melt or nuclei proliferation through dendrites fracture is poor even with the external pulsed magnetic field.

Especially, because the oxide film and melt surface turbulence in the pouring process are common in the practical experiment and production conditions, the primary and dominant nuclei formation mechanism in the bulk melt must consider the interface heterogeneous nucleation and dynamic nuclei dispersion from the oxide film at the melt surface.

4.3 Interface heterogeneous nucleation and dynamic dispersion: originate from contact interface

In the practical experimental and engineering conditions, another possible interface for heterogeneous nucleation is the contact interface, including the pouring system and mold wall. In order to investigate the interface heterogeneous nucleation and dynamic dispersion from the contact interface, we analyzed the experiment 2. By periodically penetrating an alumina tube into the melt surface, the necessary undercooling and nucleation base position is provided [25, 26] at the contact interface. Therefore, massive nuclei are produced on the surface of the alumina tube and detached from the alumina surface, which refines the solidification grains. Because the nuclei dispersion is not effectively dispersed, the refined grain zones with distinct different sizes are produced.

Thus, the primary and dominant nuclei formation mechanism in the bulk melt should consider the interface heterogeneous nucleation and dynamic nuclei dispersion from the contact interface.

4.4 The survival of the dispersed nuclei

After the nuclei origin and disperse in the superheated bulk melt, the nuclei may grow to the final solid grains or may be remelted by the superheat of the inter melt. It is generally misunderstood that the melt pouring temperature will influence the melt undercooling, and the low pouring superheat corresponds with the large melt undercooling in solidification and higher nucleation density. In fact, the pouring superheat will influence the survival ability of dispersed nuclei and finally influence the solidified grain size.

According to our experiments, it is clear the interface heterogeneous nucleation and dynamic dispersion play a primary and dominant role in nuclei density contribution. Compared with traditional nucleation mechanism, there are two important and different opinions in our nucleation mechanism. First, heterogeneous nucleation happens at the special interface and nuclei will dynamically disperse into the bulk melt, while the nucleation does not happen in the bulk melt as bulk melt is undercooled, because once the bulk melt is undercooled, crystal growth from the existing nuclei at the interface needs smaller undercooling than nucleation. There is no chance for nucleation to compete with crystal growth. Second, the nucleation, dispersion, remelt, survival and growth happen from the pouring time to the time when the bulk melt solidifies. The nucleation, dispersion, remelt and survival process happen mainly when the bulk melt is with the superheat.

It is easily be misunderstood about my results that fluid motion introduced into the aluminum melt through electromagnetics or mild mechanical agitation is the reason for grain refinement. The traditional understanding that fluid motion influences the nucleation is based on the assumption that the nucleation is decided by the undercooling of the melt and that the undercooling is influenced by the fluid motion. While my mechanism is that the nucleation mainly happens at some specific positions and that the fluid motion carries the detached nuclei into the part or whole melt. Although the fluid motion is favorable for grain refinement, the contribution role of fluid motion is rather different.

## 5. Conclusions

The nuclei in the bulk melt primitively origin at some specific interface, and subsequently dynamically disperse into the bulk melt with solid movement and melt flow. The final solidified grain size is decided not only by above process but also by the survival and growth of dispersed nuclei. The whole nucleation process is highly dynamic, evolving and variable with actual experimental and production condition.

**Acknowledgement**

This research is financially supported by the Youth Innovation Promotion Association of China Academy of Sciences. This research is supported by Chinese National S&T Major Project (2009AZ04014-081).

# Figures

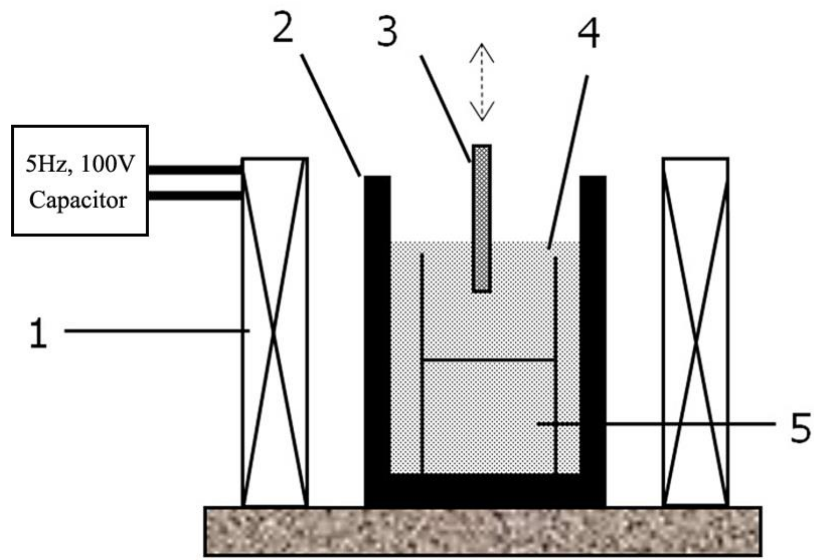

Figure 1 Sketch for the experimental apparatus. 1. Pulsed magnetic field, 2. Graphite mold, 3. Penetrating alumina tube, 4 Stainless steel sieve, 5 Melt.

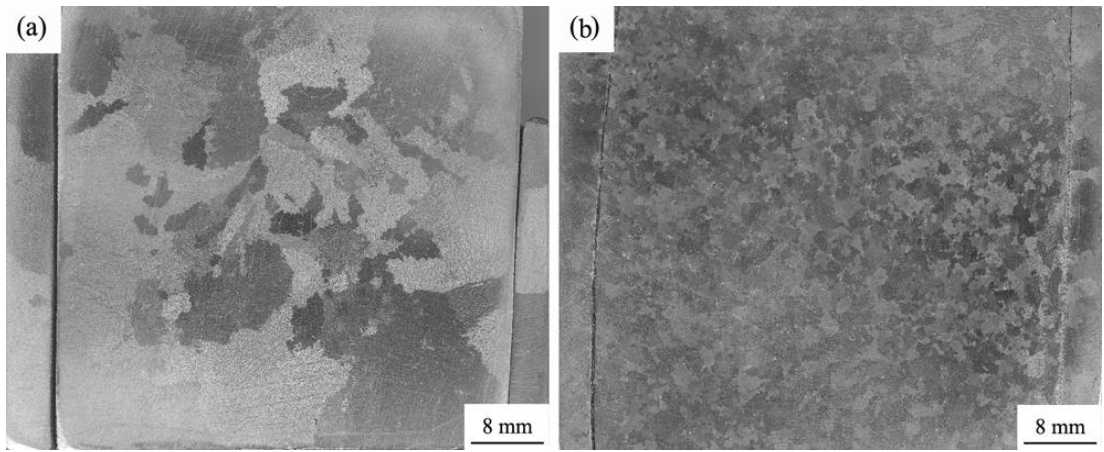

Figure 2 The solidified macrostructures of Al-5%Cu ingots. (a) with stainless steel sieves, but without pulsed magnetic; (b) with stainless steel sieves and pulsed magnetic.

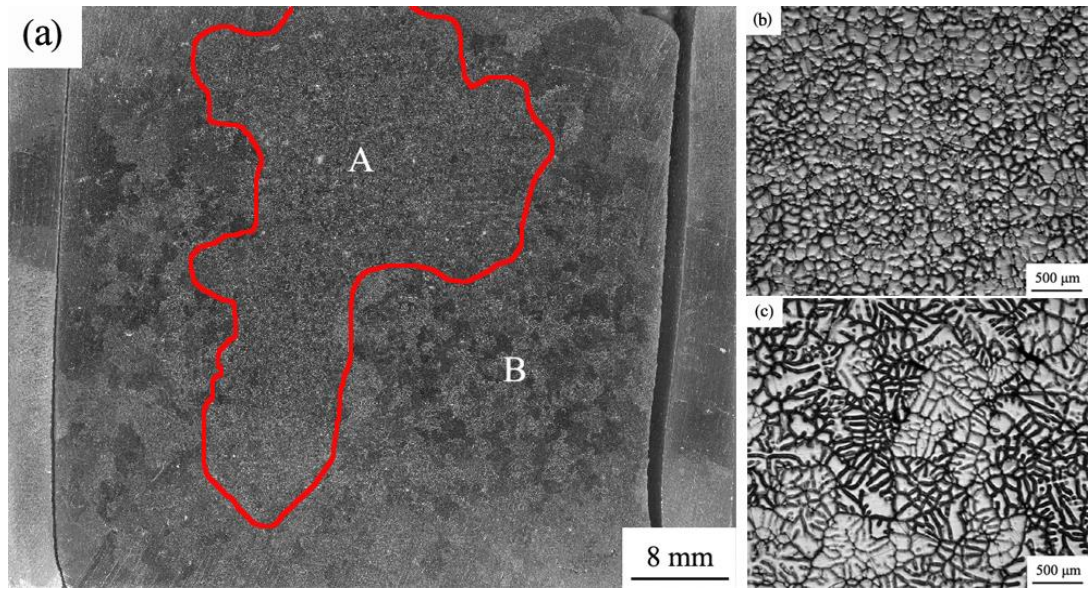

Figure 3 The solidified macrostructures of Al-5%Cu ingot with stainless steel sieves and periodic alumina tube penetration, but without pulsed magnetic field. (a) the macrostructure; (b) the finer grains in A zone; (c) the coarser grains in B zone

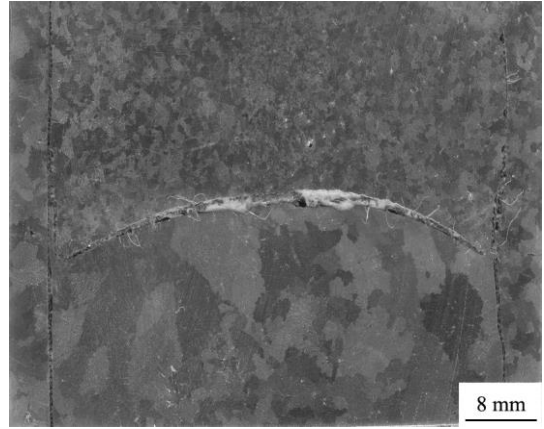

Figure 4 The solidified macrostructure with stainless steel sieves, pulsed magnetic field and periodic alumina tube penetration.